\documentclass[%
 reprint,
 superscriptaddress,
 amsmath,amssymb,
 aps,
]{revtex4-1}

\usepackage[usenames, dvipsnames]{color}
\usepackage{graphicx}
\usepackage{dcolumn}
\usepackage[utf8]{inputenc}
\usepackage{bm}
\usepackage{siunitx}
\usepackage{float}

\begin{document}

\preprint{APS/123-QED}

\title{Photon Assisted Tunneling of Zero Modes in a Majorana Wire}

\author{David M.T.~van~Zanten}
\thanks{These authors contributed equally to this work}
\affiliation{Center for Quantum Devices, Niels Bohr Institute, University of Copenhagen, and Microsoft Quantum Lab Copenhagen, Universitetsparken 5, 2100 Copenhagen, Denmark}

\author{Deividas~Sabonis}
\thanks{These authors contributed equally to this work}
\affiliation{Center for Quantum Devices, Niels Bohr Institute, University of Copenhagen, and Microsoft Quantum Lab Copenhagen, Universitetsparken 5, 2100 Copenhagen, Denmark}

\author{Judith~Suter}
\thanks{These authors contributed equally to this work}
\affiliation{Center for Quantum Devices, Niels Bohr Institute, University of Copenhagen, and Microsoft Quantum Lab Copenhagen, Universitetsparken 5, 2100 Copenhagen, Denmark}

\author{Jukka~I.~V\"{a}yrynen}
\affiliation{Microsoft Quantum, Station Q, University of California, Santa Barbara, California 93106-6105, USA}

\author{Torsten~Karzig}
\affiliation{Microsoft Quantum, Station Q, University of California, Santa Barbara, California 93106-6105, USA}

\author{Dmitry~I.~Pikulin}
\affiliation{Microsoft Quantum, Station Q, University of California, Santa Barbara, California 93106-6105, USA}

\author{Eoin~C.~T.~O'Farrell}
\affiliation{Center for Quantum Devices, Niels Bohr Institute, University of Copenhagen, and Microsoft Quantum Lab Copenhagen, Universitetsparken 5, 2100 Copenhagen, Denmark}

\author{Davydas~Razmadze}
\affiliation{Center for Quantum Devices, Niels Bohr Institute, University of Copenhagen, and Microsoft Quantum Lab Copenhagen, Universitetsparken 5, 2100 Copenhagen, Denmark}

\author{Karl~D.~Petersson}
\affiliation{Center for Quantum Devices, Niels Bohr Institute, University of Copenhagen, and Microsoft Quantum Lab Copenhagen, Universitetsparken 5, 2100 Copenhagen, Denmark}

\author{Peter~Krogstrup}
\affiliation{Microsoft Quantum Materials Lab, and Center for Quantum Devices, Niels Bohr Institute, University of Copenhagen, Kanalvej 7, 2800 Kongens Lyngby, Denmark}

\author{Charles~M.~Marcus}
\email[email: ]{marcus@nbi.ku.dk}
\affiliation{Center for Quantum Devices, Niels Bohr Institute, University of Copenhagen, and Microsoft Quantum Lab Copenhagen, Universitetsparken 5, 2100 Copenhagen, Denmark}

\date{\today} 
\maketitle

\twocolumngrid

\textbf{
Hybrid nanowires with proximity-induced superconductivity in the topological regime host Majorana zero modes (MZMs) at their ends, and networks of such structures can produce topologically protected qubits. In a double-island geometry where each segment hosts a pair of MZMs, inter-pair coupling mixes the charge parity of the islands and opens an energy gap between the even and odd charge states at the inter-island charge degeneracy. Here, we report on the spectroscopic measurement of such an energy gap in an InAs/Al double-island device by tracking the position of the microwave-induced quasiparticle (qp) transitions using a radio-frequency (rf) charge sensor. In zero magnetic field, photon assisted tunneling (PAT) of Cooper pairs gives rise to resonant lines in the 2\textit{e}-2\textit{e} periodic charge stability diagram. In the presence of a magnetic field aligned along the nanowire, resonance lines are observed parallel to the inter-island charge degeneracy of the 1\textit{e}-1\textit{e} periodic charge stability diagram, where the 1\textit{e} periodicity results from a zero-energy subgap state that emerges in magnetic field. Resonant lines in the charge stability diagram indicate coherent photon assisted tunneling of single-electron states, changing the parity of the two islands. The dependence of resonant frequency on detuning indicates a sizable (GHz-scale) hybridization of zero modes across the junction separating islands. 
}

\begin{figure*}
    \includegraphics[scale=0.8]{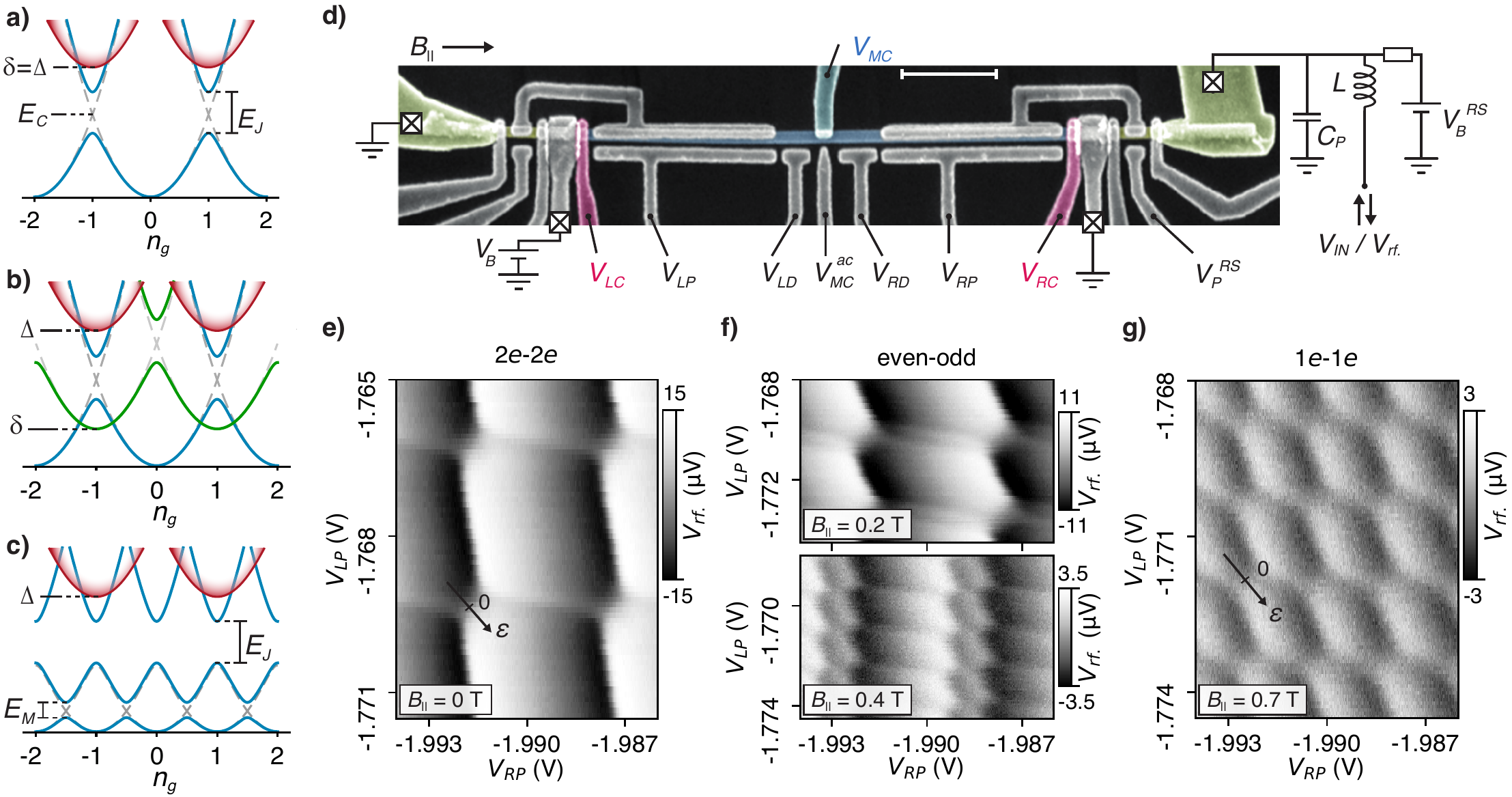}
    \caption{\textbf{Charge states of semiconductor-superconductor double-island device for different magnetic field values} 
    \textbf{a-c}, Charge state energy diagrams of superconducting island ($\Delta > E_C$) for different values of the lowest quasiparticle state $\delta$ i.e. $\delta = \Delta$ (\textbf{a}), $\Delta > \delta > 0 $ (\textbf{b}) and $\delta = 0 $ (\textbf{c}).
    \textbf{d}, False colored scanning electron micrograph of the device with the gates labeled and the magnetic field orientation indicated. Superconducting islands, formed by the tunnel gates $V_{LC}$, $V_{MC}$ and $V_{RC}$ are capacitively coupled to normal SET charge sensors embedded in \textit{RLC} resonant circuits. The size of the scalebar is 1~$\mu$m.
    \textbf{e}, Superconducting double island charge stability diagram at $B_{||}$ = 0 and $V_{MC} = -109.5$~mV as a function of $V_{RP}$ and $V_{LP}$ as measured by the rf-signal reflection ($V_{rf.}$) of the right SET charge sensor. The direction of the detuning axis $\varepsilon$ is indicated by the arrow. 
    \textbf{f}, Similar to \textbf{e} but at $B_{||}$ = 0.2~T (top panel) and $B_{||}$ = 0.4~T (middle panel). 
    \textbf{g}, Similar to \textbf{e} but at $B_{||}$ = 0.7~T. The direction of the detuning axis $\varepsilon$ in the 1\textit{e} regime is indicated by the black arrow.}
    \label{fig:fig1}
\end{figure*}

In a mesoscopic superconducting island, the presence of a single unpaired electron comes at an energy cost $\delta$ determined by the lowest quasiparticle state. As a consequence, the odd-parity charge states are elevated in energy. The total energy of a charge state with $N$ electrons is $E_N = E_C(N-n_g)^2 + \delta\, \mathrm{mod}(N,2)$, where $E_C$ is the single-electron charging energy and $n_g$ is the gate-induced charge. For $\delta > E_C$ (Fig.~\ref{fig:fig1}a), the island ground states are of even charge parity only, and the island occupation changes in steps of 2\textit{e} as $n_g$ is changed \cite{Lafarge1993nat}. When the island is weakly coupled to a superconducting reservoir the coherent exchange of Cooper pairs, characterized by Josephson energy $E_J$, mixes ground states of equal parity, resulting in an anti-crossing (blue lines in Fig.~\ref{fig:fig1}a). This superposition of even-parity charge states is the basis of conventional superconducting qubits \cite{Nakamura1999}. 

In a topological superconducting island, non-overlapping Majorana zero modes comprise a zero-energy qp state denoted a Majorana bound state (MBS). Since $\delta = 0$ in this case, the island charge occupation changes with regular steps of width 1\textit{e}. Overlap between MZMs within the island leads to $\delta \ne 0$. Consequently, the width of Coulomb valleys with even parity will either increase or decrease depending on the sign of $\delta$ \cite{Hutzen2012}, as recently demonstrated experimentally \cite{VanVeen2018,Albrecht2016,Shen2018}. In contrast to the intra-island coupling of MZMs, the wave function overlap with an exterior MZM, characterized by $E_M$, leaves $\delta$ unchanged but mixes states of different parity (see Fig.~\ref{fig:fig1}c). The ground states of the island remain 1\textit{e}-periodic, but with an anticrossing between even and odd parity states centered at half-integer values of $n_g$. The hybridization of MZMs reported here, which is key to the operation of topologically protected qubits \cite{Aasen2016, Karzig2017}, has not been demonstrated previously to our knowledge. 

In this Article, we investigate photon-assisted tunneling between coupled hybrid superconducting islands \cite{Chang2015,Krogstrup2015}. The gate-controlled double-island geometry allows independent tuning of island densities and couplings between islands and from each island to an adjacent normal-metal lead. Figure~\ref{fig:fig1}d shows a scanning electron micrograph of the device. Aluminum is selectively removed at the nanowire ends and below three barrier regions whose conductances are tuned using voltages $V_{LC}$, $V_{RC}$ and $V_{MC}$ on adjacent electrostatic gates. Coupling between islands is controlled by $V_{MC}$, while the islands are isolated from normal contacts by low-transmission tunnel junctions controlled by $V_{LC}$ and $V_{RC}$. The device is either grounded or biased with voltage $V_{B}$. Charge of the right island is measured using a capacitively coupled segment of the same nanowire (see Fig.~1a) configured as single electron transistors (SET)\cite{Lafarge1993nat} and embedded in an on-chip \textit{RLC} resonant circuit \cite{Schoelkopf1998, Razmadze2018}. 

Charge stability diagrams of the double island at several axial magnetic fields are shown in Fig.~1e-g. Sharp horizontal (vertical) transitions indicate changes in occupancy of the  left (right) island only, mediated by charge transfer from adjacent normal metallic leads. Less sharp transitions are also visible along the diagonal detuning axis (marked by $\varepsilon$ in Fig.~1e and 1g), where charge is transferred between the two islands with fixed total charge. For these diagonal, total-charge-preserving transitions, the double-island system can be considered as a single island (Fig.~\ref{fig:fig1}a-c) in a differential charge basis (see Supplementary~Note~B) $n_g$ along the $\varepsilon$ axis, with effective charging energy $E_C = E_C^L + E_C^R - E_C^m$, where $E_C^L$, $E_C^R$, $E_C^m$ are the left-island, right-island, and mutual charging energies.

Note in Fig.~1e-g that the periodicity of charge transitions changes with magnetic field. The regular transitions along the horizontal and vertical axis at $B_{||}$ = 0 (Fig.~\ref{fig:fig1}e) split at intermediate field values, respectively $B_{||} =$ 180~mT and $B_{||} \simeq 250$~mT, resulting in a pattern of wide and narrow Coulomb valleys (Fig.~\ref{fig:fig1}f). By $B_{||}$ = 0.7~T, a regular pattern of transitions (Fig.~\ref{fig:fig1}g) is again observed, now with half the period compared to  $B_{||}$ = 0. This evolution of charge states with magnetic field is consistent with the transition from 2\textit{e}-periodic ground states to 1\textit{e}-periodic ground states illustrated in Fig.~\ref{fig:fig1}a-c. 

\begin{figure*}
  \includegraphics[scale=0.8]{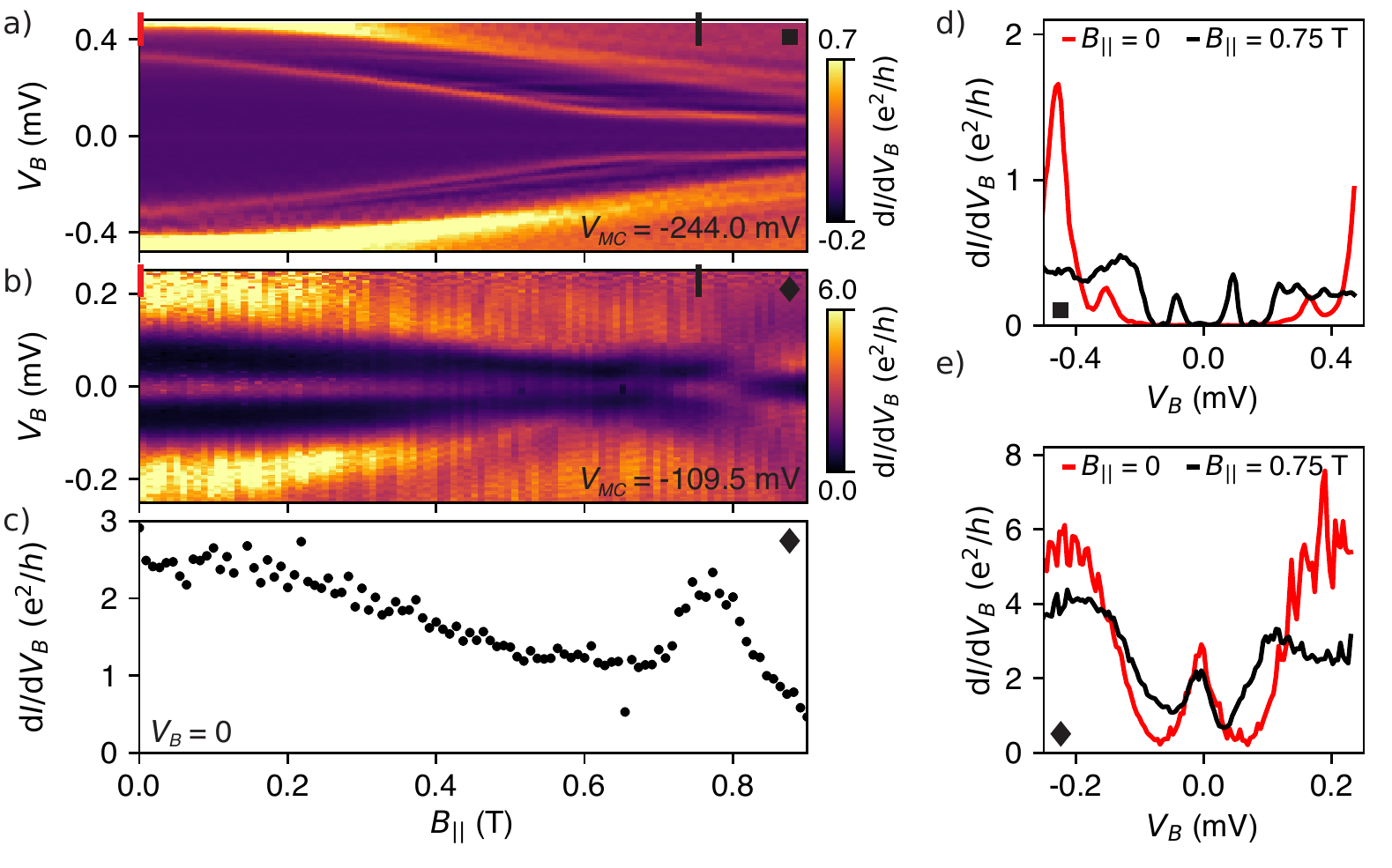}
  \caption{\textbf{Characterization of tunnel conductance through the middle tunnel-junction as a function of magnetic field parallel to the nanowire.} 
  \textbf{a}, Differential conductance $\mathrm{d}I/\mathrm{d}V_B$ as a function of $V_B$ and $B_{||}$ measured in low transmission regime ($V_{MC} = -244$~mV), showing the persistence of hard-gapped density of states in high $B_{||}$ (linecuts at $B_{||} = 0$ and $B_{||} = 0.75$~T shown in \textbf{d}). 
  \textbf{b}, Same as \textbf{a} but measured in moderate transmission regime ($V_{MC} = -109.5$~mV) showing the appearance of a soft-gap behavior (linecuts at $B_{||} = 0$ and $B_{||} = 0.75$~T shown in \textbf{e}) and finite zero-bias conductance. A horizontal linecut at $V_B = 0$ is plotted in \textbf{c}.}
  \label{fig:fig2}
\end{figure*}

At $B_{||}$ = 0, both islands have 2\textit{e}-periodic ground states and the transitions in charge occupation along $\varepsilon$ are broadened by the Josephson coupling $E_J$ between the islands. In contrast, vertical and horizontal total-charge transitions are only thermally broadened. Increasing $B_{||}$ reduces the energy of extended states in the quasiparticle continuum \cite{Oreg2010} and intrinsic (localized) subgap states (i.e. Andreev bound states in the islands) \cite{Huang2018} by the Zeeman energy. This results in a decrease of $\delta_{L/R}$ leading to the emergence of odd-parity ground states around the degeneracies of 2\textit{e} ground states for $\delta_{L/R} < E_C^{L/R} - E_C^m$. 

Further increase of $B_{||}$ yields 1\textit{e}-periodic ground states, in this device occurring first in the left island (for $B_{||} > 450$~mT) followed by right island (for $B_{||} > 600$~mT) as $\delta_{L/R}$ approaches zero. In addition to the change in periodicity in $n_g$, we find an offset $\Delta n_g^L \sim \Delta n_g^R \sim 0.5$ between 2\textit{e} and 1\textit{e} degeneracy points (see Supplementary~Figure ~\ref{fig:sup4}). This offset excludes an incoherent equal-weight mixture of different total parity states as the cause of the 2\textit{e} to 1\textit{e} transition \cite{VanVeen2018,Joyez94}. The smallest value of $B_{||}$ yielding 1\textit{e}-periodic ground states, together with $\Delta = 235$~$\mu$eV (obtained from conductance data shown in Fig.~\ref{fig:fig2}), gives an upper bound on the effective $g$-factor of the lowest energy sub-gap-state in the island, which is $2\Delta / (\mu_B B_{||}) =$~17 and 13.5 for the left and right islands, respectively. 

We note that a transition from 2\textit{e} to 1\textit{e} periodic charge states can also be caused by the reduction of the trivial (induced) superconducting gap \cite{Lutchyn2010,Lafarge1993prl}, in contrast to the picture of a discrete state at zero energy. To distinguish these possibilities, we examine the local density of states (DOS) of the nanowire by measuring the differential conductance between islands as function of voltage bias, $V_B$, with the outer barriers fully opened by applying positive gate voltage to the outer gates, $V_{LC}=V_{RC}=+0.5$~V. Other relevant gates, $V_{LP}$, $V_{LD}$, $V_{RD}$ and $V_{RP}$, are left unchanged. As the tunnel junction couples two superconducting segments, the conductance may contain contributions from supercurrent \cite{Ingold1992,Holst1994,Pekola2010} and multiple Andreev reflection (MAR) which obscure the underlying DOS \cite{Zhang2018}. However, these contributions are suppressed at low-transmission ($\mathrm{d}I/\mathrm{d}V \ll \mathrm{e}^2/h$), and tunnel-conductance of the middle junction accurately reflects in the DOS adjacent wire segments. 

Figure~\ref{fig:fig2}a shows tunneling conductance as a function of $V_B$ and $B_{||}$ at $V_{MC} = -244$~mV. The high-bias ($V_B = 0.5$~mV) conductance is $\sim 0.5$ $\mathrm{e}^2/h$ indicating that the measurement is taken in the low-transmission regime. The induced superconducting gaps in the wire-segments ($\Delta_L$ and $\Delta_R$) clearly result in the suppression of qp tunneling for $\mathrm{e}|V_B| < \Delta_L + \Delta_R$. A line cut at $B_{||} = 0$ (red curve in Fig.~\ref{fig:fig2}d) shows strong resonances in the conductance at $V_B = \pm 470$~$\mu$V (referred to as gap resonance) which we associate with tunneling between the qp continua. From this we estimate $\Delta_L + \Delta_R = 470$~$\mu$eV. This is consistent with a value of twice the induced gap, as reported by other works on InAs/Al nanowires \cite{Albrecht2016,Deng2016,Vaitiekenas2018,Mannila2018}, from which we conclude  $\Delta_R \sim \Delta_L \sim \Delta$ at $B_{||}=0$~T. The same line cut also shows a single sub-gap resonance at $V_B = \pm 320$~$\mu$V, which we attributed to qp tunneling between a sub-gap state at $E = \pm85$~$\mu$eV in one wire-segment and the qp continuum in the other. 

With increasing $B_{||}$ this simple picture develops into a more complex set of resonances. The gap resonance initially decreases in a slightly non-linear manner and splits into several resonances. This is consistent with the Zeeman splitting of the intrinsic subgap states, i.e. Andreev bound states (ABS) in the nanowire \cite{Huang2018}. From $B_{||} > 0.4$~T the lowest gap resonance continues to decrease linearly with a slope of $g_{\rm eff} = 7.1$, which is significantly smaller than upper bound on the $g_{\rm eff}$ extracted before. The field dependent DOS on both sides of the tunnel junction makes it hard to address the origin of each of these resonances and consequently to identify the presence of MZMs. Instead we focus on the bias-spectroscopy measured in the magnetic field regime where we observe a regular  1\textit{e}-1\textit{e} charge stability diagram (see Fig.~\ref{fig:fig2}d). A line cut of Fig.~\ref{fig:fig2}a taken at $B_{||} = 0.75$~T (black curve in Fig.~\ref{fig:fig2}d) shows the persistence of a hard-gapped DOS in both wire-segments. 

Having demonstrated hard-gapped DOS in both wire-segments at finite parallel magnetic field, we now characterize conductance in the regime of Fig.~\ref{fig:fig1}, where the middle barrier is more open  ($V_{MC} = -109.5$~mV). We show the conductance as a function of $V_B$ and $B_{||}$ in Fig.~\ref{fig:fig2}b. This regime, where the above-gap conductance exceeds 1 $\mathrm{}{e}^2/h$, a sizable sub-gap conductance is observed, arising from the combination of supercurrent, multiple Andreev reflections, and activated qp transport, and therefore only qualitatively and indirectly reflect the induced gap $\Delta$. In this regime, we observe significant zero-bias conductance with a nonmonotonic amplitude with increasing  $B_{||}$, as seen in Fig.~\ref{fig:fig2}c. For $B_{||} < 0.7$~T, we attribute zero-bias conductance to the transport of Cooper pairs (CPs), that is, supercurrent. This is supported by the observation of a 2\textit{e} - 2\textit{e} or even/odd-even/odd charge stability diagrams (Fig.~\ref{fig:fig1}e and \ref{fig:fig1}f). Around $B_{||} = 0.7$~T the conductance at $V_B = 0$ suddenly increases which signals the presence of an additional conduction channel. The increase in conductance at $V_B = 0$, supported by the observed 1\textit{e}-periodic charge states (Fig.~\ref{fig:fig1}g) and $\Delta > 90$~$\mu$eV (Fig.~\ref{fig:fig2}a), is consistent with the presence of MZMs in both wire-segments. 

\begin{figure}
  \includegraphics[width=8.5cm]{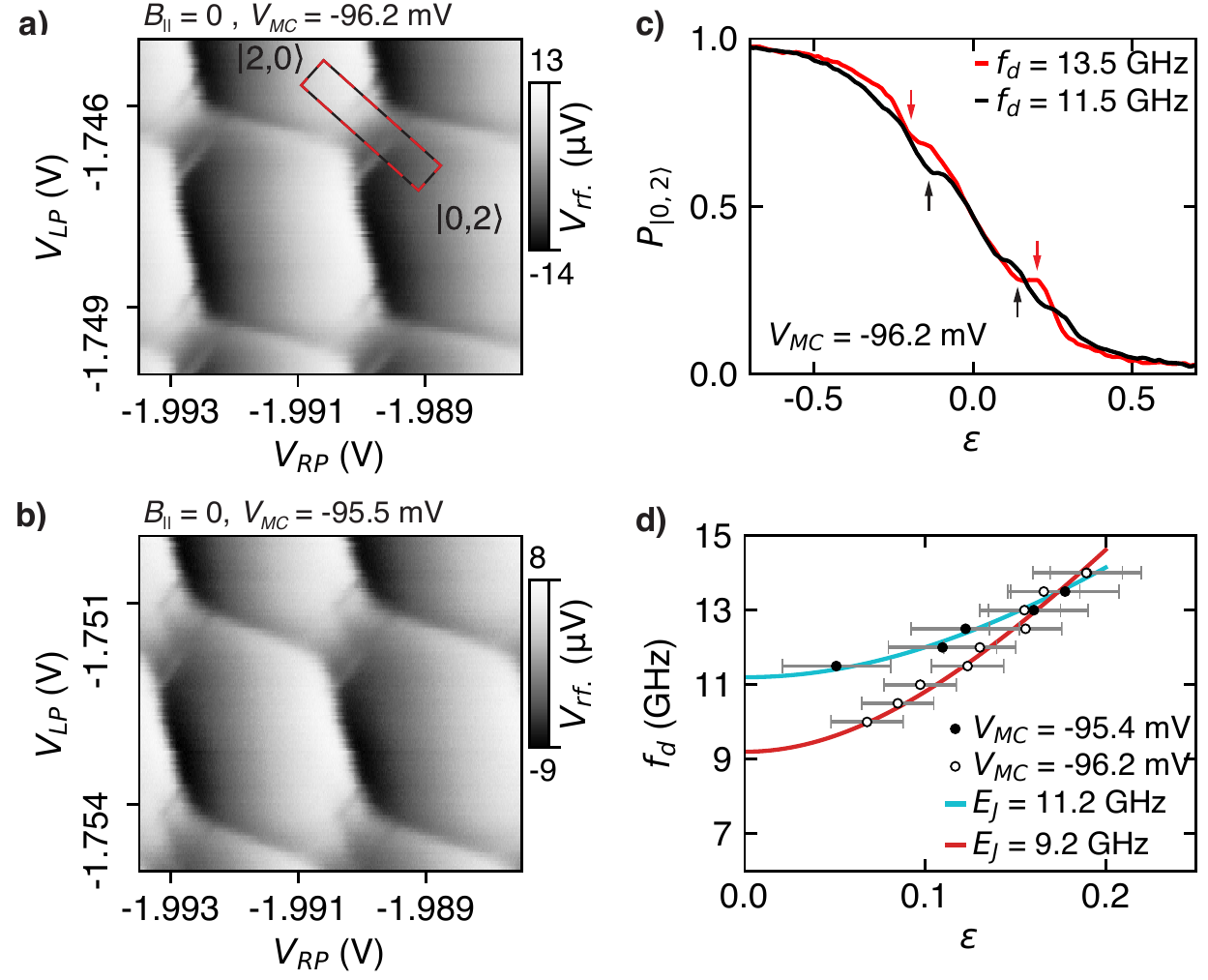}
  \caption{\textbf{Microwave spectroscopy at zero magnetic field.} 
  \textbf{a}, Superconducting double-island charge stability diagram at $V_{MC} = -95.5$~mV as a function of $V_{RP}$ and $V_{LP}$ as measured by the rf-signal reflection of the right SET charge sensor. A continuous microwave drive signal ($f_d = 13.5$ GHz) was applied to $V_{RP}$ during the measurement, which makes resonance lines appear parallel and around the charge degeneracy points. 
  \textbf{b}, Similar to \textbf{a} but with different hybridization between islands ($V_{MC} = -96.2$~mV). 
  \textbf{c}, Cut along the diagonal (detuning axis $\varepsilon$) of data set in \textbf{b} at two different microwave frequencies: $f_d = 13.5$~GHz (red) and $f_d = 11.5$~GHz (black). Features appearing due to microwave driving are indicated by red and black arrows. Trace is normalized to state (0,2) occupation probability. 
  \textbf{d}, Energy dispersion as a function of $\varepsilon$ for two different $V_{MC}$ values. The errorbars are based on the FWHM of the resonance feature. Data is fitted to a parabolic energy band approximation with $E_{J}/h = 11.2$~GHz (magenta) and $E_{J}/h = 9.2$~GHz (red).}
  \label{fig:fig3}
\end{figure}

Having characterized the double-island system, we now apply microwaves through a high bandwidth line connected to the right plunger gate, labeled $V_{RP}$. Figure~\ref{fig:fig3}a shows the zero-field charge stability diagram when the device is continuously irradiated at frequency $f_d = 13.5$~GHz. For $hf_d \geq E_J$, resonance lines appear parallel to the diagonal charge-transition line, a signature of photon assisted tunneling (PAT) through the middle tunnel barrier \cite{Lehnert2003,Duty2004,Lambert2017}. The fact that the resonance appears as a line rather than a step indicates that PAT occurs between discrete quantum states, as expected for PAT between superconductors mediated by the exchange of CPs. The coupling between islands can be controlled by the tunnel gate $V_{MC}$. A stronger coupling results in more rounding around the triple points, as shown in Fig.~\ref{fig:fig3}b. 

The peak position along the detuning axis is set by the condition $hf_d = E_{ex}$, where $E_{ex}$ is the energy difference between the ground state and the first excited state. In Fig.~\ref{fig:fig3}c we show a (normalized) line-cut of Fig.~\ref{fig:fig3}a along $\varepsilon$ (red), together with a line-cut generated from data taken using $f_d = 11.5$~GHz (black). The PAT resonances, indicated by red and black arrow, symmetrically move away from $\varepsilon = 0$, with increasing frequency. The full width half maximum (FWHM) of the peaks is about 5~$\mu$eV, but increases with the microwave power \cite{Petersson2010,Lehnert2003}. In supplementary material we investigate the FWHM at different excitation powers such that we can extract the inhomogeneous broadening by extrapolation of the data to zero power (Supplementary~Figure~\ref{fig:sup1}). We find an intrinsic broadening of $\sim 4$~$\mu$eV which implies an upper bound on inhomogeneous coherence time, $T_{2}^{*} < 1$~ns, for the 2\textit{e} superconducting charge qubit.

Frequency dependence of the zero-field PAT resonance is shown in Fig.~\ref{fig:fig3}d for two middle barrier settings. Each data point is extracted from separate charge stability diagrams exposed to microwaves. Fitting to the form $E_{ex} = \sqrt{16(\varepsilon E_C)^2 + E_J^2}$ yields fit parameters $E_C/h = 10.8$~GHz, $E_J/h = 11.2$~GHz (cyan curve) and $E_C/h = 14.2$~GHz, $E_J/h = 9.2$~GHz (red curve). The change in $E_C$ at different middle barriers is presumably dominated by changes in $E_C^m$, as the horizontal and vertical size of the hexagons remain constant when changing $V_{MC}$. These curves demonstrates the sensitivity of $E_J$ to middle barrier voltage, $V_{MC}$. A different regime with $E_J/h = 16$~GHz is shown in the supplementary data (Supplementary~Figure~\ref{fig:sup5}).

\begin{figure}[t]
  \includegraphics[width=8.5cm]{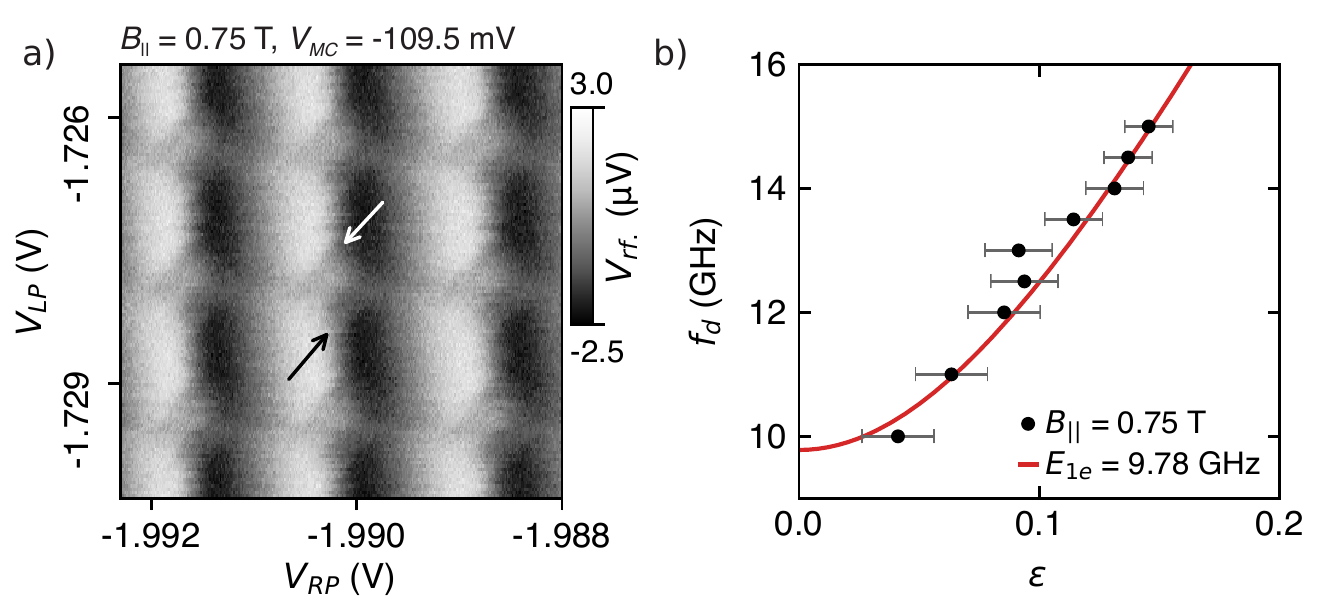}
  \caption{\textbf{Microwave spectroscopy at $\mathbf{\textit{B}_{||}} = 0.75$ T.} 
  \textbf{a}, Superconducting double-island charge stability diagram at $V_{MC} = -109.5$~mV as a function of $V_{RP}$ and $V_{LP}$. A continuous microwave signal ($f_d = 14$ GHz) was applied to $V_{RP}$ during the measurement, which makes resonance lines appear parallel and around the charge degeneracy points. 
  \textbf{b}, Energy dispersion as a function of $\varepsilon$ at $V_{MC} = -109.5$~mV. The errorbars are based on the FWHM of the resonance feature. Data is fitted to a parabolic energy band approximation and give $E_{1\textit{e}}/h = 9.9$~GHz across the tunneling barrier. }
  \label{fig:fig4}
\end{figure}

Figure~\ref{fig:fig4}a shows a charge stability diagram at $B_{||} = 0.75$~T in the same gate-voltage regime as Fig.~\ref{fig:fig3} at applied microwave frequency $f_d = 13.5$~GHz. Similar to Fig.~\ref{fig:fig1}a and \ref{fig:fig1}c, the charge stability diagram is 1\textit{e}-periodic at this field. Resonance lines are again observed parallel to the diagonal charge-transition line, now centered around the 1\textit{e}-1\textit{e} degeneracies. The observation of 1\textit{e}-periodic PAT resonance lines suggests exchange of quasiparticles between discrete zero-energy states, which could be either zero-energy ABS or MZMs. To confirm the nature of the resonance lines, we trace their position along the $\varepsilon$-axis as a function of the microwave drive frequency $f_d$ (black dots in Fig.~\ref{fig:fig4}b). The monotonous increase of peak position with $f_d$ is consistent with the dispersion of 1\textit{e} charge states. 

The coherent exchange of quasiparticles couples states of different parity, giving an avoided crossing at $\varepsilon = 0$ with characteristic energy splitting $E_{1\textit{e}}$. Within a Majorana interpretation of the discrete subgap states that coherently couple, this splitting is the Majorana coupling $E_M$ across the junction. However, because we cannot unambiguously distinguish MZMs from zero-energy ABSs in this set up, we instead refer to the characteristic splitting at $\varepsilon = 0$ as $E_{1\textit{e}}$. The expected form for the anticrossing for $E_J = 0$ and $E_{1\textit{e}}\leq E_C$, is $E_{ex}^{1\textit{e}} \approx \sqrt{E_{1\textit{e}}^2 + 4(\varepsilon E_C)^2}$. Fits to the data using $E_{ex}^{1\textit{e}} = hf_d$ yield the fit parameters $E_{1\textit{e}}/h = 9.8 \pm 0.18$~GHz and $E_C/h = 38.8 \pm 1.2$~GHz. 

Though we observe a significant decrease in Cooper-pair conductance with increasing magnetic field (Fig.~\ref{fig:fig2}c), it is unlikely that $E_J$ goes to zero at $B_{||} = 0.75$~T. To account for a non-zero Josephson coupling we numerically diagonalize the Hamiltonian and calculate the energy difference between the lowest two eigenstates as a function of detuning for fixed Josephson coupling. This procedure allows us to to fit the data of Fig.~\ref{fig:fig4}b and we obtain a little improved agreement (Supplementary~Figure~\ref{fig:sup3}) with parameters $E_{1\textit{e}}/h = 9.6$~GHz, $E_J/h = 0.48$~GHz and $E_C/h = 40.5 \pm 1.2$~GHz. A $\chi^2$ analysis as a function of $E_{1\textit{e}}$ and $E_J$ for fixed $E_C$ shows that in this regime $E_C > E_{1\textit{e}}$, the fit is nearly insensitive to the value of $E_J$ (Supplementary~Figure~\ref{fig:sup2}). 

Finally, we note that the line-width of the PAT resonance at $B_{||} = 0.75$~T is somewhat larger than at zero magnetic field. From the line-width we estimate $T_{2}^{*} \sim 100$~ps. We cannot point to a single cause, but speculate that reduction of the induced superconducting gap leads to faster decoherence. Another possibility is an increase of electronic temperature. The PAT features in general look less pronounced at $B_{||} = 0.75$~T which we attribute to the decrease of signal-to-noise ratio of the charge-sensor (64 at $B_{||} = 0$ and 11 at $B_{||} = 0.75$~T).

In conclusion, we have investigated the field-driven transition from 2\textit{e}-periodic to 1\textit{e}-periodic charge states in a hybrid InAl/Al double-island device with integrated charge sensors, then carried out a study of photon assisted tunneling in both the 2\textit{e} and 1\textit{e} regimes. Tunnel-spectroscopy of the junction between the islands shows a hard-gapped density of states for $B_{||} < 0.9$~T, whereas transport data show a clear increase in zero-bias conductance around $B_{||} =0.75$~T. Photon assisted tunneling at $B_{||} =0.75$~T revealed coherent coupling between discrete zero-energy states, consistent with coupling of Majorana zero modes across the junction. 

\section*{Methods}
\subsection*{Device fabrication}
The InAs nanowire was grown by molecular beam epitaxy with subsequent deposition of epitaxial Al (10~nm) on two of its facets. The nanowire was deposited using a micro-manipulator, onto a Si (525~$\mu$m) chip with a local Ti/Au backgate  (5~nm/35~nm) passivated by Si$_{3}$N$_{4}$ (30~nm) and HfO$_{2}$ (15~nm). The two mesoscopic superconducting islands separated by a Josephson junction were defined using the etchant Transene D (9~s, $\ang{48}$~C). The nanowire was contacted by Ti/Au (5~nm/150~nm) normal leads defined by electron beam lithography.  Subsequently, the device was coated by HfO$_{2}$ (7~nm) before evaporating Ti/Au (5~nm/ 160~nm) side and top gates under rotation of the sample stage and partially under an angle (60~nm \ang{0}, 15~nm \ang{5}, 85~nm \ang{0}).

\hfill \break
\subsection*{Measurement techniques}
Measurements were carried out in an Oxford Instruments Triton-400 dilution refrigerator with a base electron temperature of \textit{T} $\sim 50$~mK and a 6-1-1~T vector magnet. Differential conductance $g = \mathrm{d}I/\mathrm{d}V$ was measured using the ac-lockin techniques with an excitation voltage in the range 4-10~$\mu$V and excitation frequencies below 200~Hz

The charge sensor is embedded in the resonant circuit by bonding the reflectometry line to one of its leads. The sensor is configured in a Coulomb blockade regime and plunger $V_{P}$ is set such that we stay on the slope of the single peak. The sensing principle is based on the fact that charges tunnelling in and out of the double-island will modify the electrostatic environment beyond the trivial direct capacitive interaction between two gates. The action of electron or quasiparticle tunneling into and island will move the sensor island out of its most sensitive position (slope) and hence will modulate the reflected from the tank circuit signal. 

For resonant circuit excitation and demodulation we used Zurich Instruments Ultra High Frequency Lock-in amplifier \cite{UHFLI}. The excitation signal coming from the lock-in is first attenuated by 21~dB of built-in cryostat attenuators and then by another 15~dB additionally as it passes though the CPL-to-In port of the directional coupler installed below the mixing chamber.

After reflection from the tank circuit, the signal is amplified by $\sim 40$~dB using a cryogenic amplifier (Weinreb CITLF3) followed by a lock-in amplifier, where it is digitally demodulated with a demodulation time constant of 30~ns. For increasing signal to noise ratio each data point was averaged a 1000 times. To record charge stability diagrams we used a rastered gate scanning similar to \cite{Stehlik2015}. To avoid the capacitive cross talk between the $V_{RP}$ and the sensor plunger $V_{SP}$ we apply an out of phase compensating saw-tooth ramp at the same frequency on $V_{SP}$ while taking a charge stability diagram. The fast measurement setup used here is equivalent to the one presented in \cite{Razmadze2018}.

\section*{Acknowledgments}
We thank S.~Upadhyay for help with fabrication. Research is supported by Microsoft and the Danish National Research Foundation. Judith Suter acknowledges financial support from the Werner Siemens Foundation Switzerland. Peter Krogstrup acknowledge support from European Research Commission, grant number 716655. Charles Marcus acknowledges support from the Villum Foundation.

\clearpage

\appendix
\twocolumngrid

\setcounter{figure}{0} 
\renewcommand{\figurename}{Supplementary Figure}

\section*{Supplementary Data}

\begin{figure}[H]
  \includegraphics[width=8.0cm]{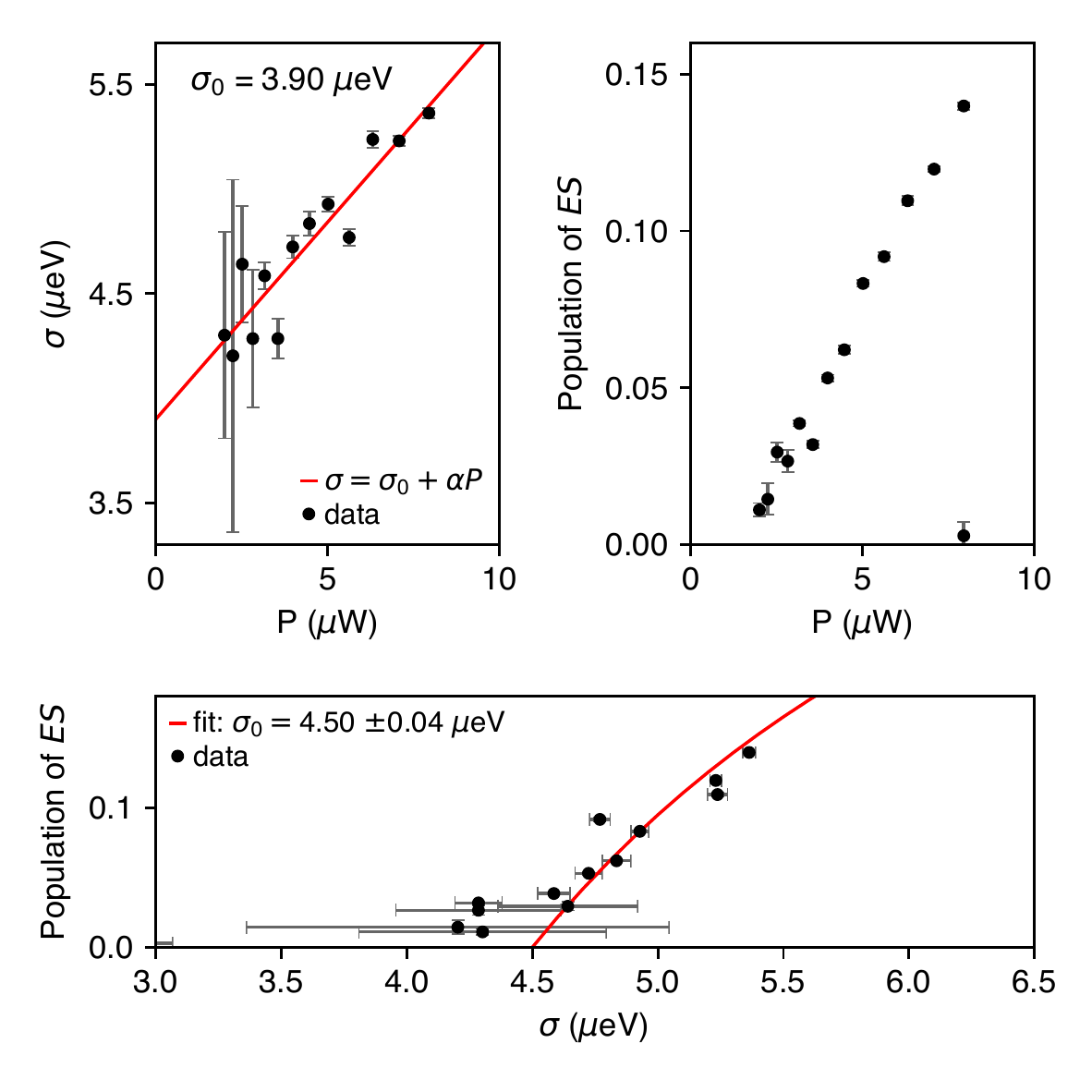}
  \caption{\textbf{Study of broadening of photon assisted tunneling resonances (PAT) with increasing microwave power.} 
  \textbf{a}, Gaussian width ($\sigma$) of the PAT resonances as a function of the microwave power (circles). Solid red line is a linear fit of the data. 
  \textbf{b}, PAT resonance peak height normalized to 2\textbf{e} charge step as a function of microwave power. 
  \textbf{c}, Parametric plot of the data in \textbf{a} and \textbf{b} (circles). Solid red line is a fit using $y = 0.5(1-\sigma_0/x)^2$ with $\sigma_0 = 4.5 \pm 0.04$~$\mu$eV.}
  \label{fig:sup1}
\end{figure}

\begin{figure}[H]
  \includegraphics[width=8.0cm]{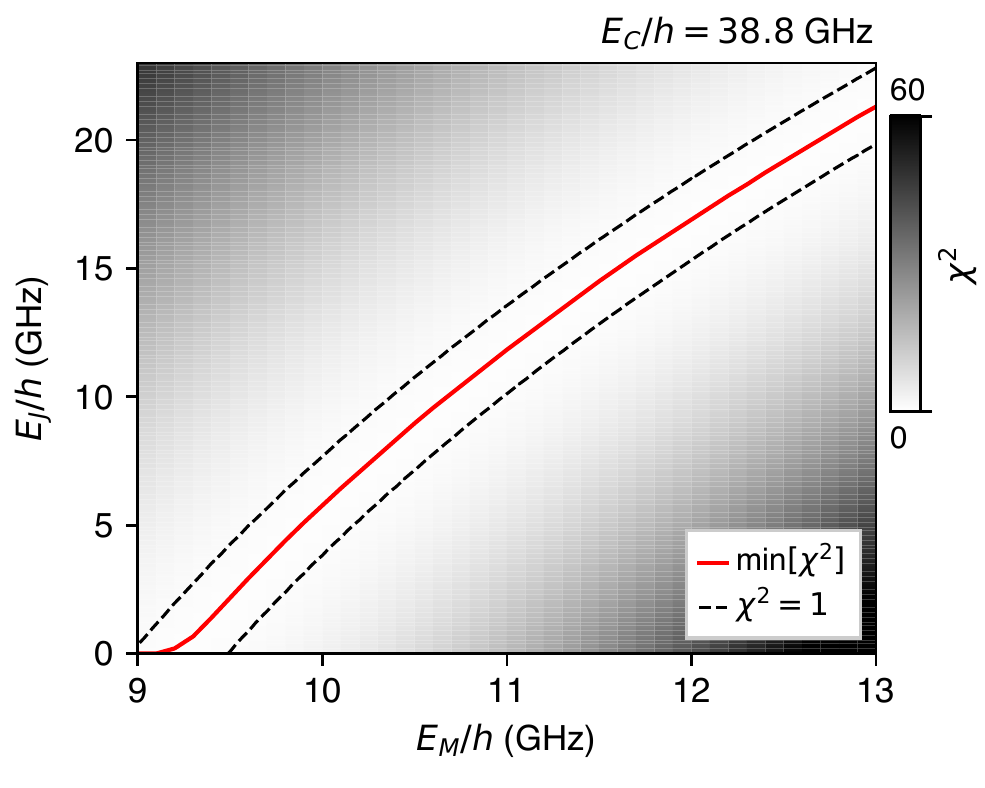}
  \caption{\textbf{$\chi^2$-analyses of 1\textit{e}-periodic charge dispersion data.} 
  Colormap of the $\chi^2$-analyses of the data presented in the Fig.~\ref{fig:fig4}b. The minimum $\chi^2$ value and the $\chi^2=1$ contour are plotted in red and black respectively.}
  \label{fig:sup2}
\end{figure}

\begin{figure}[H]
  \includegraphics[width=8.0cm]{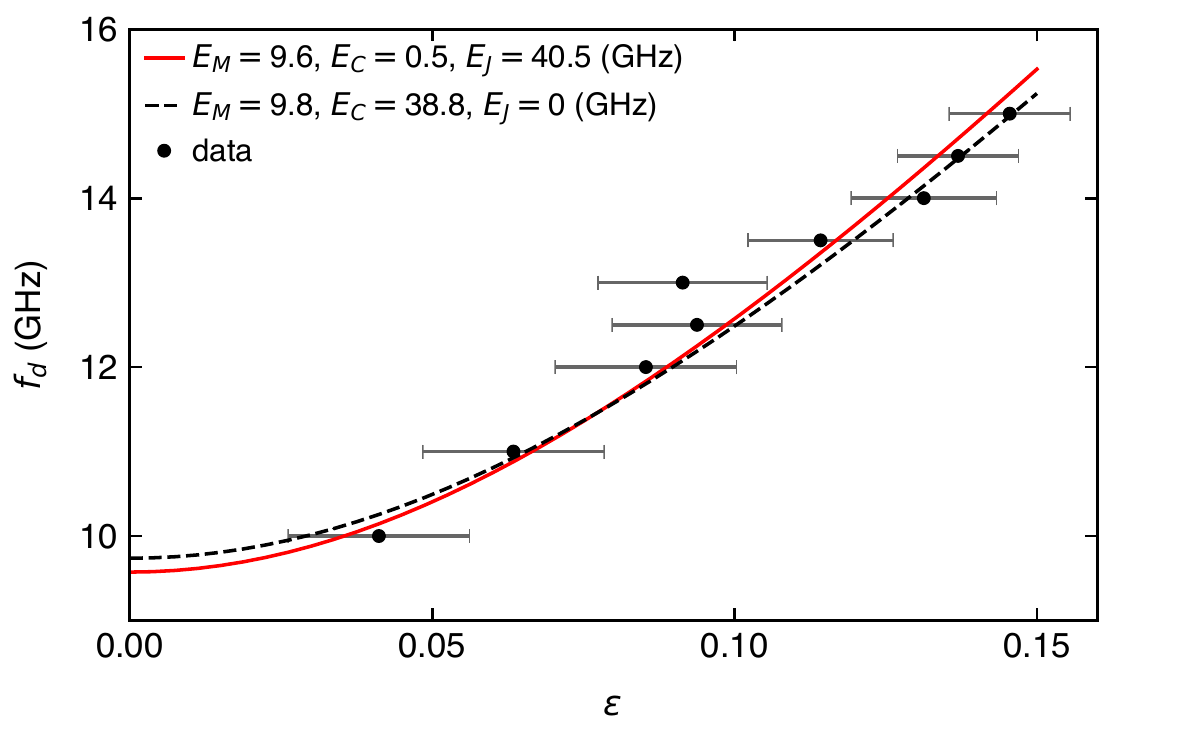}
  \caption{\textbf{Data of Fig.~\ref{fig:fig4}b (circles) together with the results of two fit models.}  
  Dashed black line is a parabolic fit assuming $E_J = 0$ ($\chi^2$ = 1.88). Solid red lines is a numerical fit taking into account $E_J$ ($\chi^2$ = 1.56).}
  \label{fig:sup3}
\end{figure}

\begin{figure}[H]
  \includegraphics[width=8.0cm]{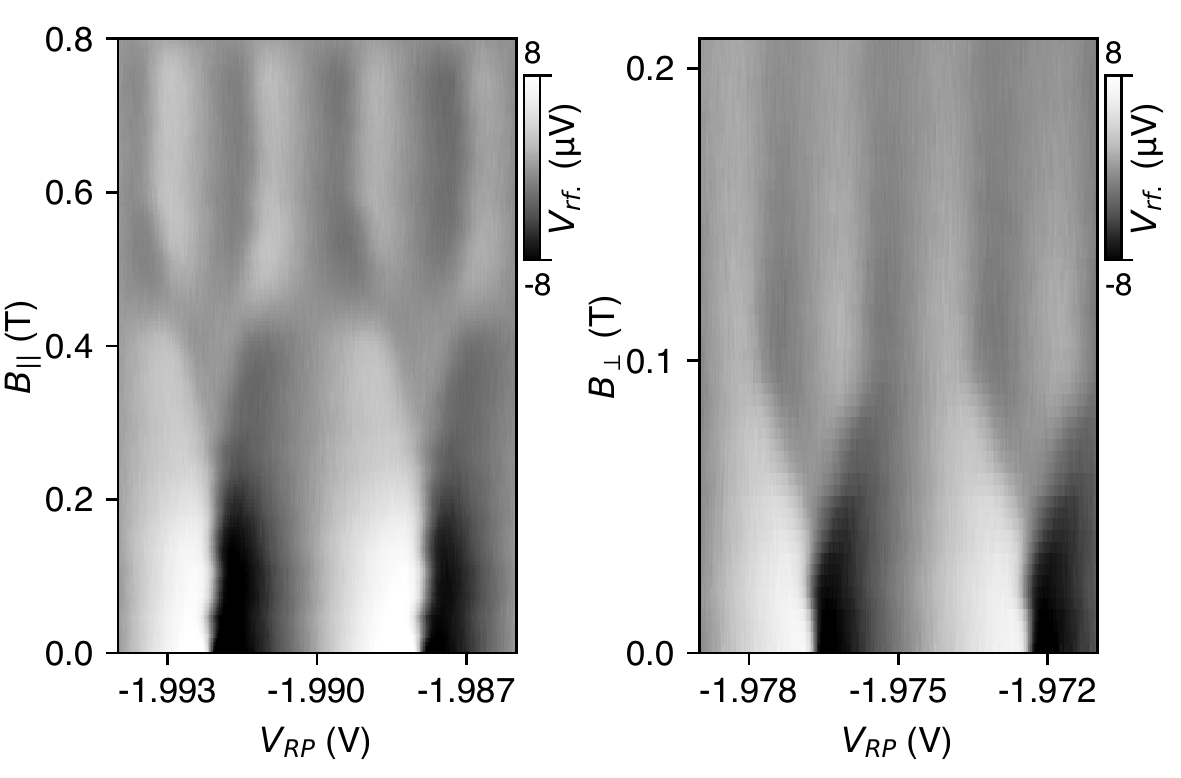}
  \caption{\textbf{Evolution from 2\textit{e} to 1\textit{e} charge steps.}
  \textbf{a}, Right charge sensor signal as a function of $V_{RP}$ and magnetic field parallel to the wire $B_{||}$ at $V_{LP} = -1.754$~V. 
  \textbf{b}, Right charge sensor signal as a function of $V_{RP}$ and magnetic field perpendicular to the sample-plane $B_\perp$ at $V_{LP} = 1.770$~V.}
  \label{fig:sup4}
\end{figure}

\begin{figure}[H]
  \includegraphics[width=8.0cm]{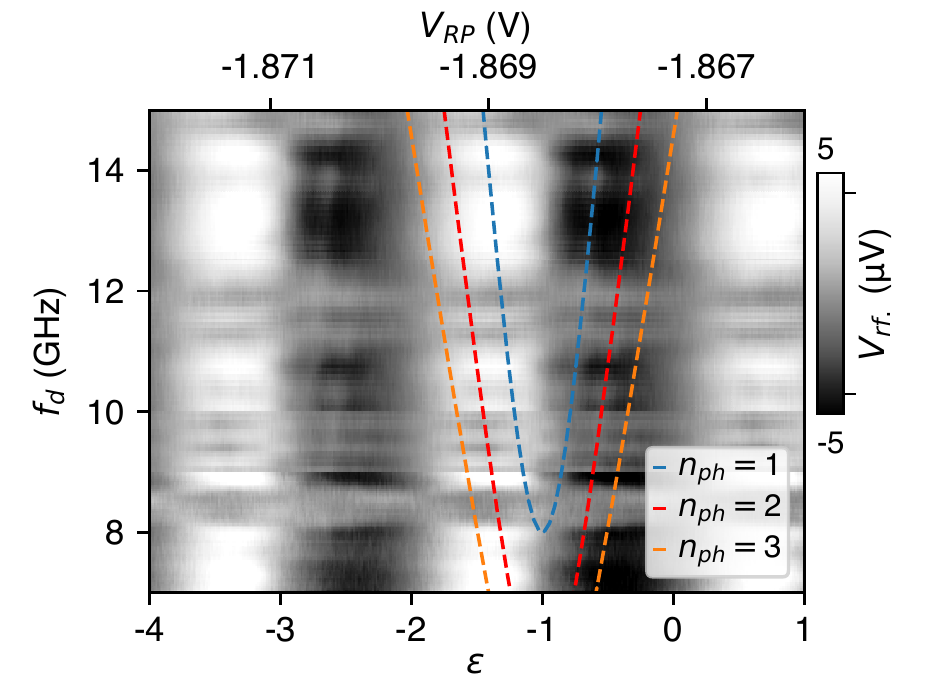}
  \caption{\textbf{2\textit{e}-periodic dispersion of multi-photon photon-assisted tunneling features.} 
  Right charge sensor signal as function of $V_{RP}$ and measured during irradiation with a microwave signal of varying frequency $f_d$ at $B_{||} = 0$. Dashed lines plot factors ($n_{ph}$) of the excited state energy (w.r.t ground state energy) i.e. $h f_d = \sqrt{16(\varepsilon E_C)^2 + E_J^2} / n_{ph}$, where $E_C/h$ = 14~GHz, $E_J/h$ = 16~GHz and $n_{ph}$ = 2 (blue), 3 (red) and 4 (orange). Note that the charge degeneracies are at $\varepsilon = $~-3, -2, -1.}
  \label{fig:sup5}
\end{figure}

\section*{Supplementary Note B: Charging Hamiltonian of the double-dot}\label{Suppl:ChargeH}

In this section, we introduce the charging Hamiltonian of the superconducting double-dot in the presence of weak 1-electron and 2-electron coherent tunneling.
We then project to a fixed total number of electrons and derive an effective Hamiltonian that describes the inter-dot transitions relevant to PAT.   
The Hamiltonian is $H=H_{c}+H_{\Delta}+H_{t}$
where 
\begin{align}
\begin{split}
H_{c} =& \sum_{i=L,R}E_{C}^{i}(N_{i}-n_{gi})^{2}+ E_{C}^{m}N_{L}N_{R}\,,
\end{split}\\
H_{\Delta} ={}&\sum_{i=L,R} \delta_{i}\frac{1}{2}(1-(-1)^{N_{i}})\,,\\
H_{t} =&{}E_{J}\cos\varphi+E_{1\textit{e}} \cos\frac{\varphi}{2}\,. \label{eq:SMHt}
\end{align}
Here $N_{i}$ is the operator of total number of electrons on the $i=L,R$ island, $n_{gi}$ is the compensated dimensionless gate charge (controlled by gate voltages $V_{LP}$ and $V_{RP}$), $E_{C}^{i}$ is the charging energy of that island, and $E_{C}^{m} < E_{C}^{L}+E_{C}^{R}$ is the mutual charging energy between the islands. 
Up to an irrelevant constant energy shift, the Hamiltonian $H_{c}$ is equivalent to the double-island Hamiltonian of Ref.~\cite{vdwRMP} after a redefinition of parameters. 
In the second line we included superconductivity by assigning an additional energy $\delta_{i}$ for the island $i$ to have an odd number of electrons.
In the presence of sub-gap states we have $\delta_{i} < \Delta_{i}$ while otherwise $\Delta_{i} = \delta_{i}$.  

In Eq.~(\ref{eq:SMHt}) above we introduced tunneling of Cooper pairs and single
electrons; 
in the charge basis, 
\begin{flalign}
\cos\frac{\varphi}{2} & =\frac{1}{2}|N_{L}+1,N_{R}-1\rangle\langle N_{L},N_{R}|+h.c.\,,\\
\cos\varphi & =\frac{1}{2}|N_{L}+2,N_{R}-2\rangle\langle N_{L},N_{R}|+h.c.
\end{flalign}

Let us write the charging Hamiltonian in terms of the differential 
and total charges, $N_{-}=\frac{1}{2}(N_{L}-N_{R})$, $N_{+}=N_{L}+N_{R}$.
The tunneling terms  in this basis  are 
\begin{flalign}
\cos\frac{\varphi}{2} & =\frac{1}{2}|N_{-}+1\rangle\langle N_{-}|+h.c.\,,\\
\cos\varphi & =\frac{1}{2}|N_{-}+2\rangle\langle N_{-}|+h.c.\,
\end{flalign}
and they commute with $N_{+}$, as does the entire Hamiltonian. 
In order to describe an isolated double-island, we project the Hamiltonian to an eigenstate of $N_{+}$. The projected charging Hamiltonian has an effective single-island form,  
\begin{equation}
H_{c}=E_{C}(N_{-}-n_{g})^{2}+const.\,,
\end{equation}
where 
\begin{flalign}
E_{C} & =E_{C}^{L}+E_{C}^{R}-E_{C}^{m}\,,\\
n_{g} & =\frac{1}{2E_{C}}\left[E_{C}^{R}(N_{+}-2n_{gR})-E_{C}^{L}(N_{+}-2n_{gL})\right]\,.
\end{flalign}
The full Hamiltonian projected to a fixed $N_{+}$ is then (ignoring
the $N_{+}$-dependent constant)
\begin{flalign}
  H_{c} & =E_{C}(N_{-}-n_{g})^{2}\,, \label{eq:SMHc} \\
  \begin{split}
      H_{\Delta} & =\delta_{L}\frac{1}{2}[1-(-1)^{N_{-}+\frac{1}{2}N_{+}}]+ \\
      & \delta_{R}\frac{1}{2}[1-(-1)^{-N_{-}+\frac{1}{2}N_{+}}]\,,
  \end{split} \label{eq:SMHdelta} \\
  \begin{split}
      H_{t} & = \frac{1}{2} E_{J} | N_{-}+2 \rangle \langle N_{-}|+ \\   
      & +\frac{1}{2} E_{1\textit{e}} |N_{-}+1\rangle \langle N_{-}|+h.c.
  \end{split}
\end{flalign}

Next, we will study the vicinity of the inter-island charge-degeneracy points in two cases  of (i) no sub-gap states on either island ($\delta_i =\Delta_i$), and (ii) a zero-energy sub-gap state on each island ($\delta_i =0$).
The two cases respectively correspond to $2e$ and $1e$ periodic charge stability diagrams  (see Fig. \ref{fig:fig1}e and \ref{fig:fig1}g in the main text). 

\subsection{$2e$ degeneracy points}\label{Suppl:Fit2e}

At zero magnetic field the ground state of the double dot has no unpaired electrons.
Then,  $N_{+}$ is an even integer and $N_{-}$ is an integer with the same parity as $N_{+}/2$. 
Let us suppose for concreteness that  $N_{+}/2$ is fixed to an even integer (the odd case is treated similarly). 
Then $N_{-}$ takes only even values. 
The inter-island charge degeneracy points occur when $n_g$ in Eq.~(\ref{eq:SMHc}) is an odd integer. 
Consider $n_{g}\approx1$ so that it is enough to only include  the states $|N_{-}\rangle =|0\rangle\,,|2\rangle $ in the low-energy description. 
Then we have a $2\times 2$ effective Hamiltonian in the charge basis, \begin{equation}
H_{\text{eff},2e}=\left(\begin{array}{cc}
E_C + 2E_{C}[n_{g}-1] & \frac{1}{2}E_{J}\\
\frac{1}{2}E_{J} & E_C-2E_{C}[n_{g}-1]
\end{array}\right)\,.
\end{equation}
The eigenenergies are  $E_{\pm}=E_{C}\pm\frac{1}{2}\sqrt{E_{J}^{2}+16(E_{C} [n_{g}-1])^{2}}$.
Upon microwave irradiation, the single-photon resonance frequency is therefore $\omega=E_{+}-E_{-}=\sqrt{E_{J}^{2}+16(E_{C} \varepsilon )^{2}}$, where $\varepsilon = n_{g}-1 $.

\subsection{$1\textit{e}$ degeneracy points}\label{Suppl:Fit1e}

Let us consider a strong magnetic field so that we have a zero-energy sub-gap state on each island. Then we can set $\delta_{L,R}=0$ in the Hamiltonian, Eq.~(\ref{eq:SMHdelta}), since an  unpaired electron can be added to  the sub-gap state without pairing energy cost. 
The degeneracy is then at half-integer values of $n_{g}$ and the distance between degeneracy points is $1$. 
For concreteness, let us take $n_{g}\approx1/2$ and only consider charge states $N_{-}=0,1$.
Then we have a $2\times 2$ effective Hamiltonian in the
basis $\{|0\rangle,\,|1\rangle\}$, 
\begin{equation}
H_{\text{eff},1e} =\left(\!\begin{array}{cc}
-E_{C}[\frac{1}{2}-n_{g}] & \frac{1}{2}E_{1\textit{e}}\\
\frac{1}{2}E_{1\textit{e}} & E_{C}[\frac{1}{2}-n_{g}]
\end{array}\!\right)+\frac{1}{4}E_{C}\,,
\end{equation}
and energies $E_{\pm}=\frac{1}{4}E_{C}\pm\frac{1}{2}E_{1\textit{e}}\sqrt{1+4(\frac{E_{C}}{E_{1\textit{e}}})^{2}[\frac{1}{2}-n_{g}]^{2}}.$
The single-photon resonance frequency is
\begin{equation}
\omega=\sqrt{E_{1\textit{e}}^{2}+4(E_{C}\varepsilon)^{2}}\,,
\end{equation}
where $\varepsilon = n_{g}-\frac{1}{2}$.

\end{document}